\documentclass[prd,twocolumn,aps,superscriptaddress,nofootinbib,showpacs]{revtex4-1}
\usepackage{amsmath}
\usepackage{amssymb}
\usepackage{graphicx}
\usepackage{color}
\usepackage{epsfig}
\usepackage{dcolumn}
\usepackage{float}
\usepackage{lipsum}
\usepackage{soul}
\usepackage{bm}
\usepackage{times}
\usepackage{hyperref}
\hypersetup{
  colorlinks=true,
  citecolor=blue,
  linkcolor=blue,
  urlcolor=blue}

\usepackage{epsfig}
\usepackage{dcolumn}
\usepackage{morefloats}
\def\be{\begin{equation}}
\def\ee{\end{equation}}
\def\bea{\begin{eqnarray}}
\def\eea{\end{eqnarray}}
\def\gsim{\ \rlap{\raise 2pt\hbox{$>$}}{\lower 2pt \hbox{$\sim$}}\ }
\def\lsim{\ \rlap{\raise 2pt\hbox{$<$}}{\lower 2pt \hbox{$\sim$}}\ }
\def\dslash{\kern-4pt \not{\hbox{\kern-2pt $\partial$}}}
\def\pslash{\not{\hbox{\kern-2pt p}}}

\def\th23{{$\theta_{23}$}}

\begin{document}

\renewcommand{\arraystretch}{2}
\DeclareGraphicsExtensions{.eps,.ps}


\title{A New Sensitivity Goal for Neutrino-less Double Beta Decay Experiments }



  \author{Vishnudath K. N.}
\email[Email Address: ]{vishnudath@prl.res.in}
\affiliation{Physical Research Laboratory, Ahmedabad - 380009, India}
\affiliation{Discipline of Physics, Indian Institute of Technology, Gandhinagar - 382355, India}

\author{Sandhya Choubey}
\email[Email Address: ]{sandhya@hri.res.in}
\affiliation{Harish-Chandra Research Institute, HBNI,
Chhatnag Road, Jhunsi, Allahabad 211 019, India}

 \author{Srubabati Goswami}
\email[Email Address: ]{sruba@prl.res.in}
\affiliation{Physical Research Laboratory, Ahmedabad - 380009, India}

\begin{abstract}

We study the implications of the Dark-LMA solution to the solar neutrino 
problem for neutrino-less double beta decay ($0\nu\beta\beta$ ). We show that 
while  the predictions  for the effective mass  governing $0\nu\beta\beta$  
remains unchanged for the inverted mass scheme, that for normal ordering 
becomes higher for the Dark-LMA parameter space and moves into the 
``desert region'' between the two. 
This sets a new goal for sensitvity reach for the
next generation experiments 
if no signal is found for the inverted ordering by the future search 
programmes.

\end{abstract}
\maketitle


\textbf{\textit{Introduction}} : 
Fermions can be of two types - Dirac or Majorana.  A Majorana particle is a self-conjugate fermion. All known fermions other than the neutrino are 
Dirac particles. Neutrinos, being neutral, are the only known particle that can possibly be a Majorana fermion. Hence, the question whether neutrinos are Dirac or Majorana particle is one of the most fundamental questions in physics.
The most straightforward way to probe the Majorana nature of neutrinos is 
through neutrino-less double beta decay ($0\nu\beta\beta$). 
While beta decay involves the decay of a radioactive nucleus into a 
daughter nucleus along with an electron and an electron-type antineutrino, 
neutrino-less double beta decay is a rare process in which a nucleus of 
charge Z and mass number A decays 
into a daughter, producing two electrons and no neutrinos
: $(A,Z)\to (A,Z+2)+2e^-$ \cite{Furry:1939qr}.  
A positive signal of this   
will be a definite confirmation of the existence of lepton number violating 
Majorana mass term for the neutrinos \cite{Schechter:1980gr}.
Such a term requires transcending beyond
the Standard  Model of particle physics  and could also be related to the 
observed preponderance of matter over antimatter which is essential for 
our existence. 
Thus it is not surprising that 
searches for $0\nu\beta\beta$ have been on-going for the past several 
decades \cite{Barabash:2011mf}. 
While no undisputed positive signal has been seen in any of the experiments 
so-far, a lower limit (90\% C.L.) on the $0\nu\beta\beta$ lifetime of 
$T_{1/2} (^{136}{Xe})  > 1.5 \times 10^{25}$ years has been obtained from KamLAND-Zen \cite{KamLAND-Zen:2016pfg}, $T_{1/2} (^{76}{Ge})  > 8 \times 10^{25}$ years from GERDA
\cite {Agostini:2018tnm} 
and  $T_{1/2} (^{130}{Te}) > 1.5 \times 10^{25}$ years from combined results of 
CURCINO and CUORE \cite{Alduino:2017ehq}. 
In this work we assume that $0\nu\beta\beta$ is driven solely via a Majorana mass term for the neutrinos. 
Therefore, non-observation of $0\nu\beta\beta$ puts an upper limit on the effective neutrino mass which depends on the neutrino mass and mixing parameters. 
The effective mass depends crucially on whether the neutrino mass spectrum 
conforms to normal ordering (NO) or inverted ordering (IO), which corresponds 
to whether the third mass eigenstate is the heaviest or lightest, respectively. In addition to the neutrino mass ordering, the effective neutrino mass also depends on some of the other neutrino mass and mixing parameters - the two mass squared differences $\Delta m_{21}^2$ and $\Delta m_{31}^2$, two of the three mixing angles, {\it viz.}, $\theta_{12}$ and $\theta_{13}$ and the Majorana phases $\alpha_1$ and $\alpha_2$. By allowing these parameters to vary in their current $3\sigma$ allowed range, one obtains two bands of predicted values for the effective mass for IO and NO, separated by a ``desert region''. 
The effective mass corresponding to IO ($\sim$ 0.015 - 0.05 eV) 
is expected to be probed comfortably 
in the next-generation $0\nu\beta\beta$ experiments which include  
LEGEND, GERDA-II,MAJORANA D, CUPID, SNO+, KamLAND2-Zen, nEXO, NEXT 1.5K, PANDAX III 1k , SuperNEMO etc. \cite{DellOro:2016tmg}.  
While lowering the sensitivity of these experiments to be able to probe 
the effective mass for the NO case  is going to be challenging, 
it is possible to make some inroads into this region as well albeit with a lower probability \cite{Agostini:2017jim}. Many of these planned experiments will be 
capable of probing the ``desert region'' 
i.e. territories  $\lsim$ 0.01 eV even if they fall short of testing significant parts of the NO band \cite{Agostini:2017jim,Kharusi:2018eqi,Myslik:2018vts,Artusa:2014wnl}.

In this letter we show, for the first time, the impact of the so-called 
Dark-LMA (DLMA) \cite{Miranda:2004nb,Escrihuela:2009up,Farzan:2017xzy} solution to the 
solar neutrino problem on $0\nu\beta\beta$. 
The standard LMA solution corresponds to standard neutrino 
oscillations with $\Delta m_{21}^2 \simeq 7.5\times 10^{-5}$ eV$^2$ and 
$\sin^2\theta_{12} \simeq 0.3 $, and satisfies the solar neutrino data at 
high significance. The DLMA solution appears as a nearly-degenerate 
solution to the solar neutrino problem for $\Delta m_{21}^2 \simeq  
7.5\times 10^{-5}$ eV$^2$ and $\textrm{sin}^2\theta_{12} \simeq 0.7 $, 
once we allow for 
the existence of non-standard neutrino interactions (NSIs) in addition to 
standard oscillations. The KamLAND experiment is unable to break this 
degeneracy since it observes neutrino oscillations in vacuum which 
depends on $\sin^22\theta_{12}$ and hence is same for both LMA and 
DLMA solutions \footnote{Combining KamLAND and neutrino neutral current scattering experiments like CHARM to lift this degeneracy has been discussed in \cite{Escrihuela:2009up}.}. The occurrence of the DLMA solution can also adversely affect the determination of mass ordering in beam based neutrino oscillation experiments in presence of NSI \cite{Bakhti:2014pva,Coloma:2016gei,Deepthi:2016erc}. We will show that while the IO band for the effective 
mass in $0\nu\beta\beta$ experiments remains nearly same for LMA and DLMA 
solutions, the NO band gets shifted upwards for DLMA into the desert region mentioned above. As a result this may make it possible for the next-generation
experiments to start probing $0\nu\beta\beta$ for NO as well. 
This entails two-fold aspects: Firstly, this opens up unheralded regions of 
the effective neutrino mass to be probed by future $0\nu\beta\beta$ experiments. Secondly, this provides a way of testing the long-standing DLMA solution to the solar neutrino problem, irrespective of the value of the NSI parameters. Scattering experiments can also resolve this degeneracy by measuring the NSI parameters. For instance, in \cite{Coloma:2017egw}, combined constraints from neutrino oscillation and CHARM and NuTeV measurements were used to demonstrate that the degeneracy between the two LMA solutions can be resolved if NSI is only with the down quarks. Subsequently, the study performed in \cite{Coloma:2017ncl} included the COHERENT neutrino-nucleus scattering data and showed that the DLMA solution can be disfavored at the 3.1$\sigma$ and $3.6\sigma$ C.L. for NSI with up and down quarks, respectively. However, it is worth stressing that these bounds depend on the mass of the light mediator and it has been shown in \cite{Denton:2018xmq} that the COHERENT data excludes the DLMA solution at 95\% C.L. for light mediator mass $>48$ MeV only. The global analysis including oscillation and COHERENT data performed in \cite{Esteban:2018ppq} shows that the DLMA solution is still allowed at $3\sigma$, albeit for a smaller range of values of NSI parameters and for light mediators of mass $\gsim$ 10 MeV.

 Although the 
importance of precision determination of $\theta_{12}$  on the effective mass 
determined by $0\nu\beta\beta$ experiments have been highlighted
earlier \cite{Choubey:2005rq, Dueck:2011hu}, the ramifications
of the DLMA solution for $0\nu\beta\beta$ is being investigated
in this work for the first time.
 

\textbf{\textit{Predictions for \boldmath$0\nu\beta\beta$}}:
The half-life for $0\nu\beta\beta$ process in the standard three generation picture is given as,
\be \frac{\Gamma_{0\nu\beta\beta}}{\textrm{ln}2} = G \Big|\frac{M_\nu}{m_e}\Big|^2 m_{\beta\beta}^2 , \label{Thalf}\ee
where $G$ contains the phase space factors, $m_e$ is the electron mass and $M_\nu$ is the nuclear matrix element (NME). $m_{\beta\beta}$ is the effective neutrino mass given by,
\be m_{\beta\beta} = |U_{ei}^2 m_i|. \ee
$U$ is the unitary PMNS mixing matrix for the three active neutrinos and is given in the standard parametrization as, \be U = R_{23}  \tilde{R}_{13} R_{12} P \ee
where $R_{ij}$ are the three rotation matrices defined in terms of the corresponding mixing angles $\theta_{ij}$, with the Dirac CP-phase $\delta$ 
attached to $\tilde{R}_{13}$, and the phase matrix $P = \textrm{diag}\, (1,\,e^{i\alpha_2}, \, e^{i(\alpha_3 + \delta)})$  contains the Majorana phases. 
In this letter, we denote the DLMA solution for $\theta_{12}$ in the 
presence of NSI as $\theta_{D12}$ and the standard LMA solution as $\theta_{12}$. The $3\sigma$ ranges of these two parameters are given in Table \ref{parameters} \cite{Esteban:2018azc, Esteban:2018ppq}.

In this parametrization, the effective neutrino mass is,
\be m_{\beta\beta} \,\,=\,\, |m_1 \,c_{12}^2c_{13}^2 +m_2 \, s_{12}^2c_{13}^2e^{2i\alpha_2}+ m_3 \, s_{13}^2 e^{2i\alpha_3}|, \ee
where $c_{ij} = \textrm{cos}\theta_{ij}$ and $s_{ij} = \textrm{sin}\theta_{ij}$.
 $|m_{\beta\beta}| $ depends on whether the neutrino mass states follow normal or inverted ordering or they are quasi-degenerate.\\
 
 Normal ordering (NO) :  $m_1 < m_2 << m_3 $ with
\be m_2 = \sqrt{m_1^2\,+ \Delta m_{sol}^2} \,\,\, ; \,\,\,m_3 = \sqrt{m_1^2\,+ \Delta m_{sol}^2 +\Delta m_{atm}^2 } \ee

 Inverted ordering (IO)  :   $m_3 << m_1 \approx m_2 $ with
\be m_1 = \sqrt{m_3^2\,+ \Delta m_{atm}^2} \,\,\, ; \,\,\,m_2 = \sqrt{m_3^2\,+ \Delta m_{sol}^2 +\Delta m_{atm}^2 } \ee

 Quasi-degenerate (QD) : $m_1 \approx m_2 \approx m_3 >> \sqrt{\Delta m_{atm}^2} $\\
 
\begin{figure}

    \includegraphics[scale=0.625]{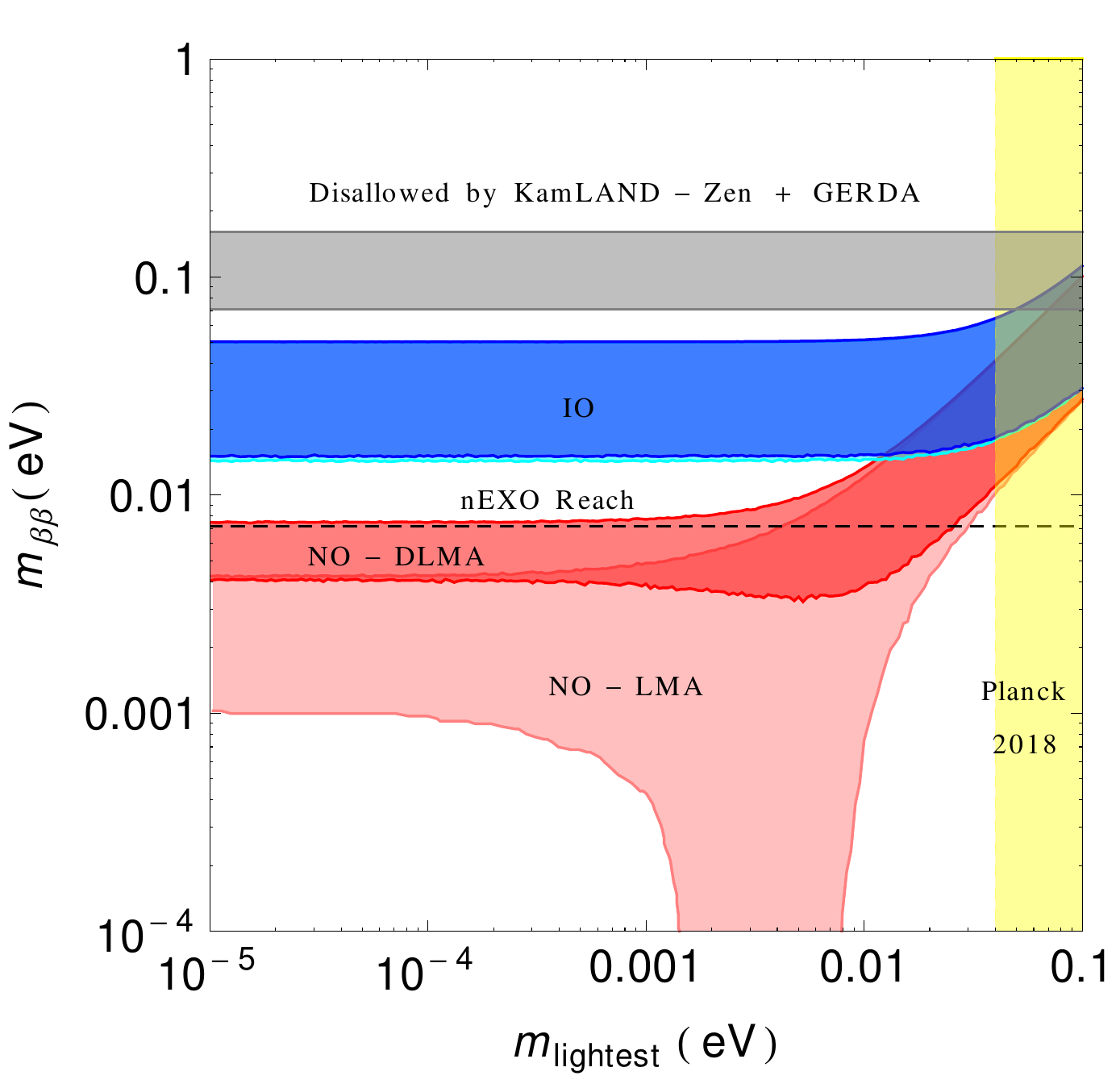}
    \caption{  The effective neutrino mass $m_{\beta \beta}$ for $0\nu\beta\beta$ as a function of the lightest neutrino mass for both NO and IO. The pink region is for NO with the standard solution for $\theta_{12}$ and the red band is for NO with $\theta_{D12}$. For the IO case(the blue band), $m_{\beta\beta}$ remains the same for the DLMA solution. See text for details.} \label{0nbb}
\end{figure}

 Here, $\Delta m_{sol}^2 = m_2^2-m_1^2$ and $\Delta m_{atm}^2 = m_3^2-m_2^2 ~(m_1^2-m_3^2) $ for NO (IO). 
Fig. \ref{0nbb} shows $m_{\beta \beta}$ as a function of the 
lightest neutrino mass for both NO and IO.
The pink region is for NO with the standard solution for $\theta_{12}$ and 
the red band is for NO with $\theta_{D12}$, corresponding to the DLMA 
solution. The dark blue band is for IO with the standard $\theta_{12}$ 
value and the cyan band (which overlaps with the blue band) is for IO with 
$\theta_{D12}$. The gray band ($0.071-0.161$ eV) corresponds to the current upper limit from 
combined results of GERDA and KamLAND-Zen experiments. The region above this is 
disallowed. The range corresponds to the NME uncertainty \cite{Engel:2016xgb,Agostini:2018tnm,Kotila:2012zza}. 
The black dashed line represents the future $3\sigma$ sensitivity of the 
nEXO experiment : $T_{1/2} = 5.7 \times 10^{27}$ years \cite{Kharusi:2018eqi}, which, for the highest value of NME, translates to $m_{\beta\beta} = 0.007$ eV. This can probe a small part of the NO region with the LMA solution for $m_{lightest} \gsim 0.005$ eV, whereas the upper edge of the DLMA region can be probed even for small values of $m_{lightest}$. The yellow region is disfavored by the 
cosmological constraints on the sum of the light neutrino masses 
\cite{Aghanim:2018eyx}. In obtaining this plot, all the oscillation parameters are varied in their 3$\sigma$ ranges \cite{Esteban:2018azc} and the Majorana phases are varied from 0 to $\pi$.
  
From the figure, we can see that for NO, $m_{\beta\beta}$ for the 
DLMA solution is higher than that for the standard LMA solution, 
shifting into the gap between IO and NO. The effect is more pronounced for 
lower values of $m_{lightest}$.
There is some overlap in the predictions between the maximum value of 
$m_{\beta\beta}$ for the LMA with the minimum value of this for the DLMA 
solution, which increases as $m_{lightest}$ increases. 
One noteworthy feature is the absence of the cancellation region for the 
DLMA solution. 
For IO, the predicted values of $m_{\beta\beta}$  remain the same for LMA and
DLMA solutions. Since the predictions of $m_{\beta\beta}$ for NO with LMA and IO with DLMA are well separated, the generalized hierarchy degeneracy \cite{Coloma:2016gei} is not present.

The behavior of $m_{\beta\beta}$ can be understood by considering the limiting cases for different mass schemes.

\underline{Inverted Ordering} : In this case, for very small values of $m_3$ such that $m_3<< \sqrt{\Delta m_{atm}^2}$, $m_2\approx m_1 \approx \sqrt{\Delta m_{atm}^2}$, the effective mass is given as, $$ {m_{\beta\beta}}_{IO} ~ \approx ~ \sqrt{\Delta m_{atm}^2}(|\,c_{12}^2c_{13}^2 + \, s_{12}^2c_{13}^2e^{2i\alpha_2}|). $$
In this region,  $m_{\beta\beta}$ is independent of $m_3$ and is bounded from above and below by a maximum and minimum value given by \cite{Pascoli:2002xq},
$$  {m_{\beta\beta}}_{IOmax}= | c_{13}^2\sqrt{\Delta m_{atm}^2} ~| \,\,\,\,\,\,\,\,\,\,\,\,\,\, (\alpha_2 = 0,\pi) , $$ $$ {m_{\beta\beta}}_{IOmin}= |c_{13}^2 ~\textrm{cos}2\theta_{12}~ \sqrt{\Delta m_{atm}^2} ~| \,\,\,\,\,\,\,\,\,\, (\alpha_2 = \pi/2)  .$$

The maximum value is independent of $\theta_{12}$ while for the minimum value, we can see from Table \ref{parameters}, that the $3\sigma$ range for $|\textrm{cos}2\theta_{12}|$ is the same for both LMA and DLMA solutions. This explains why the prediction for $m_{\beta\beta}$ is the same for both the cases in this region.

\begin{table}[ht]
 $$
 \begin{array}{|c|c|c|c|c|c|}
 \hline {\mbox {} }& \textrm{sin}^2\theta_{12} & \textrm{sin}^2\theta_{D12} & \textrm{cos}2\theta_{12}    &   \textrm{cos}2\theta_{D12}   &   \textrm{sin}^2\theta_{13}  \\
 \hline
 
  Maximum  & 0.350  & 0.725 &  0.45   & -0.30    &    0.024 \\
  
  \hline
 
  Minimum  & 0.275  &   0.650   &  0.30   & -0.45     &   0.020  \\
   
 \hline
 \end{array}
 $$
\caption{\small{ The $3\sigma$ ranges of 
different combinations of oscillation parameters 
relevant for understanding the behavior of the effective mass in different limits. 
 }}\label{parameters}\end{table}


Now, as $m_3$ approaches $ \sim \sqrt{\Delta m_{atm}^2}$, the other 
masses can be approximated as, $m_1\approx m_2\approx \sqrt{2\Delta m_{atm}^2}$ and the effective mass becomes,
$$ {m_{\beta\beta}}_{IO} = \sqrt{\Delta m_{atm}^2}~~ |( \sqrt{2}c_{13}^2(\,c_{12}^2 + \, s_{12}^2 e^{2i\alpha_2}) + \, s_{13}^2 e^{2i\alpha_3}) |.$$ This is maximum for $\alpha_2 = \alpha_3 = 0$ and is again independent of $\theta_{12}$. Also, ${m_{\beta\beta}}_{IO}$ is minimum for $\alpha_2 = \pi/2$ and $\alpha_3 = 0 $ or $\pi/2$ depending on whether we take $\theta_{12}$ or $\theta_{D12}$. But since, $s_{13}^2$ is very small, this is almost independent of what we choose for $\alpha_3$ and effectively, the minimum of ${m_{\beta\beta}}_{IO}$ in this regime is approximated as,
$${m_{\beta\beta}}_{IO min} = \sqrt{\Delta m_{atm}^2}~~ | \sqrt{2}c_{13}^2 \textrm{cos}2\theta_{12} |,$$ which is independent of the solution for $\theta_{12}$.

\underline{Normal Ordering}:
Unlike in IO, the behavior of $m_{\beta\beta}$ is different for the LMA as well as the DLMA solutions of $\theta_{12}$. For very small values of $m_1$ such that $m_1<< m_2 \approx \sqrt{\Delta m_{sol}^2} << m_3 \approx \sqrt{\Delta m_{atm}^2} $ , $m_{\beta\beta}$ can be written as,
$$ {m_{\beta\beta}}_{N0} = \sqrt{\Delta m_{atm}^2}| \sqrt{r} \, s_{12}^2c_{13}^2e^{2i\alpha_2}+ \, s_{13}^2 e^{2i\alpha_3}| ,$$
where, $ r =\lvert \frac{\Delta m_{sol}^2}{\Delta m_{atm}^2} \rvert $.
The maximum value of this corresponds to $\alpha_2 = \alpha_3 = 0,\pi$ and the minimum value corresponds to $\alpha_2 =0 $ and $ \alpha_3 = \pi/2 $. These will be higher for higher values of $\textrm{sin}^2\theta_{12}$. This explains why the prediction for $m_{\beta\beta}$ for the DLMA solution in this region is higher.

Moving on to the cancellation region, the typical values of masses are $m_1 \sim 0.005 $ eV, $m_2 \sim 0.01 $ eV and $m_3 \sim 0.05$ eV. Then, the minimum of $m_{\beta\beta}$ $(\alpha_2 =\alpha_3 =\pi/2)$ can be approximated as,
$${m_{\beta\beta}}_{min} \approx m_1|(1- 3s_{12}^2 c_{13}^2 - 11 s_{13}^2 )|.$$
For the values of $s_{12}^2$ and $s_{13}^2$ as listed in the Table \ref{parameters}, complete cancellation is possible in the LMA region. However, for $s_{12}^2$ in the DLMA region, such a cancellation is not possible because of higher values of $s_{12}^2$. 

As we increase the value of $m_1$ and reach the limit 
of partial hierarchy where 
$m_1 \approx m_2 \approx \sqrt{\Delta m_{sol}^2} << m_3 \approx \sqrt{\Delta m_{atm}^2}$, the maximum value of $m_{\beta\beta}$ is given by,
$$ {m_{\beta\beta}}_{NOmax} \approx \sqrt{\Delta m_{atm}^2 r} c_{13}^2 \,\,\,\,\,\,\,\,\,\,\,\,\,\, (\alpha_2 = \alpha_3=0) , $$ 
which is independent of $\theta_{12}$. Hence the maximum values of 
$m_{\beta\beta}$ for the two LMA solutions tend to overlap. 
In QD limit, $m_{\beta\beta}$ varies linearly with the common mass scale $m_0$ 
and both maximum and minimum values are independent of $\theta_{12}$.

  At this point it is worthwhile to note that if we assume the existence of a fourth sterile neutrino as suggested by the LSND/MiniBooNE results, then even for NO the predicted $m_{\beta\beta}$ can be in the desert region \cite{Goswami:2005ng,Barry:2011wb}. In fact, depending on the value of the mass squared difference governing the LSND/MiniBooNE oscillations, the prediction can even overlap with the IO prediction for three generation and hence, can be probed by the near future experiments.

\begin{figure}

    \includegraphics[scale=0.63]{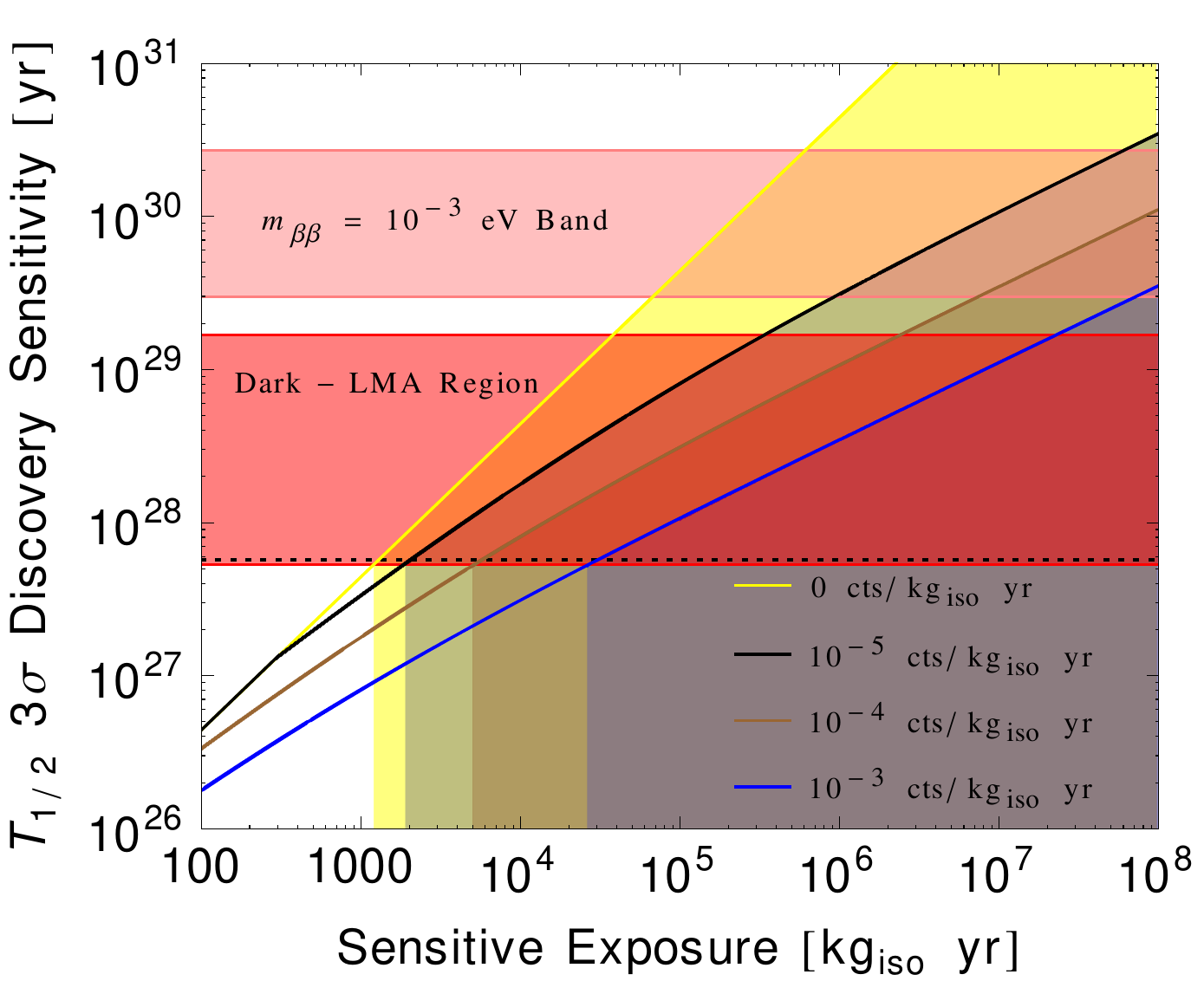}
    \caption{  ${}^{136} Xe$ discovery sensitivity as a function of 
sensitive exposure for a selection of sensitive background levels. The yellow, black, brown and blue lines correspond to different values of the sensitive background levels of $0$, $10^{-5}$, $10^{-4}$ and $10^{-3}$ $~\textrm{cts}/(\textrm{kg}_{\textrm{iso}} \textrm{ yr})$ respectively.} \label{Xe}
\end{figure}
 
\begin{table}[ht]
 $$
 \begin{array}{|c|c|c|c|}
 \hline \textrm{Isotope}& \textrm{NME}~(M_\nu) & G (10^{-15} \textrm{year}^{-1}) & T_{1/2} \, \textrm{range (years)}  \\
  
 \hline
 
  {}^{136}Xe  & 1.6-4.8 &  14.58   & 5.3\times 10^{27}-1.7\times 10^{29} \\
  
  \hline
 
 {}^{76}Ge  & 2.8-6.1 &   2.363   &  2.0 \times 10^{28} - 3.4 \times 10^{29}    \\
   
 \hline
 
  {}^{130}Te  & 1.4-6.4 &    14.22   &  4.9 \times 10^{28} - 2.2 \times 10^{29}    \\
   
 \hline
 
 \end{array}
 $$
\caption{\small{ The $T_{1/2}$ ranges corresponding to the DLMA region $m_{\beta\beta} = 0.004-0.0075$ eV for different isotopes.  The NME values \cite{Engel:2016xgb,Agostini:2018tnm} and the phase space factors \cite{Kotila:2012zza} used in the calculation are also given.
 }}\label{Isotopes}\end{table}

 
\textbf{\textit{Sensitivity in the future experiments}} : 
Here, we discuss a simple method to obtain the sensitivity of the DLMA region in the future ${}^{136}Xe$ experiments following the discussion in reference \cite{Agostini:2017jim}. The discovery sensitivity is prescribed as the value of $T_{1/2}$ for which an experiment has a $50\%$ probability of measuring a $3\sigma$ signal above the background. It is defined as,
\be  T_{1/2} = \textrm{ln}2 \frac{N_A \epsilon}{m_a S_{3\sigma}(B)} .\ee
Here, $N_A$ is the Avogadro number, $m_a$ is the atomic mass of the isotope, $B = \beta\epsilon$ is the expected background where $\epsilon$ and $\beta$ denote the sensitive exposure and background  respectively ; 
$S_{3\sigma}$ is the value for which half of the measurements would give a signal above $B$ assuming a Poisson signal and is calculated from the relation 
$$1-{CDF}_{Poisson} (C_{3\sigma}|S_{3\sigma}+B) = 50 \%.$$
$C_{3\sigma}$ denotes the number of counts for which the cumulative 
Poisson distribution with mean $B$ follows 
$CDF_{Poisson}(C_{3\sigma}|B) = 3\sigma$. 
To avoid the discrete variations that would arise in the discovery 
sensitivity if $C_{3\sigma}$ is restricted to be integer valued, 
we use the following definition of $CDF_{Poisson}$ as a continuous 
distribution in $C$ using the normalized upper incomplete gamma function,
$$ CDF_{Poisson}(C|\mu) = \frac{\Gamma(C+1,\mu)}{\Gamma(C+1)}. $$
Using the above equations, the $T_{1/2}$ discovery sensitivities of 
${}^{136}Xe$ as a function of $\epsilon$ for various values of 
$\beta$ are shown in Fig. \ref{Xe}. In this plot, the red shaded band 
corresponds to the new allowed region of $m_{\beta \beta} \sim 0.004 -0.0075$
eV  for the 
DLMA solution.
This band in $m_{\beta\beta}$ which is due to the variation of the parameters in the PMNS matrix, is converted to a band in $T_{1/2}$ using equation (\ref{Thalf}), by taking into account the NME uncertainty as given in Table \ref{Isotopes}. The pink band corresponds to $m_{\beta\beta} = 10^{-3}$ eV, which is the minimum of the NO regime for lower values of $m_{lightest}$ with the LMA solution. In Fig. \ref{Xe}, the dotted black line corresponds to the future $3\sigma$ sensitivity of nEXO, which is $T_{1/2} = 5.7 \times 10^{27}$ years \cite{Kharusi:2018eqi}. The yellow, black, brown and blue lines correspond to different values of the sensitive background levels of $0$, $10^{-5}$, $10^{-4}$ and $10^{-3}$ $~\textrm{cts}/(\textrm{kg}_{\textrm{iso}} \textrm{yr})$ respectively. From the figure, we can see that for a sensitive background level of $10^{-4}~\textrm{cts}/(\textrm{kg}_{\textrm{iso}} \textrm{yr})$, the DLMA region could be probed with a sensitive exposure greater than $\sim ~ 5000 ~\textrm{kg}_{\textrm{iso}} \textrm{yr}$. To probe the $10^{-3}$ regime shown by the dashed lines requires lower background levels and/or higher sensitive exposure. In Table \ref{Isotopes}, we have given the $T_{1/2}$ ranges corresponding to the DLMA region, $m_{\beta \beta} = 0.004 - 0.0075$ eV for three different isotopes.

\textbf{\textit{Conclusion}} : 
Searching for $0\nu\beta\beta$ process is of utmost importance 
since it can establish the Majorana nature of the neutrinos 
which implies they are their own antiparticles.  This will in-turn signify 
a lepton number violating Majorana mass term for the neutrinos, 
which may hold the key in explaining why 
neutrino masses are much smaller than the other fermion masses.
This can have profound implications for 
a deeper understanding of 
physics beyond the Standard Model of particle physics. 
So far these searches have yielded negative results and have put an 
upper bound on the 
effective mass governing $0\nu\beta\beta$.   
Assuming light Majorana neutrino exchange as the sole 
mechanism for $0\nu\beta\beta$, the predictions of effective mass for 
IO and NO are separated by a ``desert region''. The current upper bound 
is just above the IO region ($\sim 0.1$ eV ) and several future experiments with sensitivity reach $\sim 0.015$ eV
are expected to probe the IO parameter space completely. However if no positive signal 
is found in these searches then the projected sensitivity reach of these 
experiments are in the ballpark of $0.005$ eV which can 
explore only a small part of the NO region  for  lightest neutrino mass 
$\gsim$ 0.005 eV \cite{Kharusi:2018eqi}. 
The next frontier that is envisaged is $\sim 10^{-3}$ eV \cite{Penedo:2018kpc}.
In this letter, we show for the first time, that if the Dark-LMA solution 
to the solar neutrino problem is true, then the effective mass for 
NO shifts into the intermediate ``desert zone'' between NO and IO. 
Therefore, in an incremental advancement, a new goal for the $0\nu\beta\beta$ 
experiments can be to first explore this region $\sim 0.004 - 0.0075$ eV, which is possible even for very low values of the lightest neutrino mass. 
This not only defines  
a newer sensitivity goal  of future $0\nu\beta\beta$ 
experimental program for the NO scenario, 
but can also provide an independent confirmation/refutal of the Dark-LMA solution to the solar neutrino problem in presence 
of non-standard interactions. 

\section*{Acknowledgement} 
The authors would like to thank the organizers of IITB-ICTP workshop on 
neutrino physics where this idea got generated and
Frank Deppisch, K. N. Deepthi and Tanmay Kumar Poddar for useful comments.

\bibliography{nlbdk}

\end{document}